\documentclass[reprint,
showpacs,
 amsmath,amssymb,
prb,
]{revtex4-1}

\usepackage{graphicx}% Include figure files
\usepackage{dcolumn}% align table columns on decimal point
\usepackage{bm}% bold math
\usepackage{color}% bold math
\begin{document}

%\preprint{APS/123-QED}

\title{ Anomalous Proximity Effect and Theoretical Design for its Realization }
\author{Satoshi Ikegaya$^{1}$}
%\altaffiliation[Also at ]{}%Lines break automatically or can be forced with \\
\author{Yasuhiro Asano$^{1,2,4}$}
\author{Yukio Tanaka$^{3,4}$}%
%\email{\empty}
\affiliation{$^{1}$Department of Applied Physics,
Hokkaido University, Sapporo 060-8628, Japan\\
$^{2}$Center of Topological Science and Technology,
Hokkaido University, Sapporo 060-8628, Japan\\
$^{3}$Department of Applied Physics, Nagoya University, Nagoya 464-8603, Japan\\
$^{4}$Moscow Institute of Physics and Technology, 141700 Dolgoprudny, Russia}%

%\collaboration{MUSO Collaboration}%\noaffiliation
\date{\today}% It is always \today, today,
             %  but any date may be explicitly specified

\begin{abstract}
We discuss the stability of zero-energy states appearing in
a dirty normal metal attached to a 
superconducting thin film with Dresselhaus [110] spin-orbit coupling 
under an in-plane Zeeman field.
The Dresselhaus superconductor preserves an additional chiral symmetry 
and traps more than one zero-energy state at its edges.
All the zero-energy states at an edge belong to the same chirality 
in large Zeeman fields due to the effective $p$-wave pairing symmetry.
The pure chiral nature of the wave function 
enables the zero-energy states to penetrate 
the dirty normal metal while retaining their high degree of degeneracy. 
We prove the perfect Andreev reflection 
into the dirty normal metal at zero energy. 
\end{abstract}

\pacs{74.81.Fa, 74.25.F−, 74.45.+c}

\maketitle

%\tableofcontents
%*********************************************************************************
\section{Introduction}
%*********************************************************************************
%\textit{Introduction}: 
The proximity effect has been an important issue in the physics of superconductivity.
In a normal metal attached to a spin-singlet $s$-wave superconductor,
penetrating Cooper pairs form the \textsl{gap} structure in the quasiparticle 
density of states (DOS) at the Fermi level (zero energy) and modify low energy 
properties there. 
In spin-triplet $p$-wave superconductor junctions, however,
the penetrating Cooper pairs form a \textsl{zero-energy peak} in the DOS~\cite{yt05r,ya06,yt07}.
This induces various anomalous electromagnetic properties in the normal metal~\cite{yt04,ya07,ya11}.
This effect is called the anomalous proximity effect.
For instance,
a perfect Andreev reflection from a $p_x$-wave superconductor into a dirty normal 
metal causes anomalous low energy transport in the $x$-direction such as
a zero-bias conductance quantization in normal-metal/superconductor (NS) 
junctions~\cite{yt04} and
a fractional current-phase relationship in superconductor/normal-metal/superconductor (SNS)
junctions~\cite{ya06}.

Recently, these characteristic transport phenomena have been investigated 
in the context of Majorana physics~\cite{maj1,maj2} based on the topological classification of 
materials~\cite{schnyder}. In fact, as a consequence of the topologically nontrivial property 
of the wave function~\cite{chral}, the spin-triplet $p_x$-wave superconductor hosts 
more than one Majorana fermion at its edges. 
The energy dispersion of the topological edge states is flat as a function of 
the wave vector in the transverse direction (say $k_y$), which represents the high
degree of the degeneracy in the zero-energy states (ZESs). 
The anomalous proximity effect stems from the penetration of such ZESs 
into the dirty normal metal while retaining their high degree of degeneracy~\cite{yt05r,ya06,ya07}.
Theoretically, it has been unclear what symmetry protects the high degeneracy of ZESs
 and why the perfect Andreev reflection persists at zero energy.
It has been difficult to fabricate spin-triplet superconducting junctions using existing materials. 
However, the rapid progress in the Majorana physics of artificial 
superconductors~\cite{hlsc,smsc2,smsc4,smsc5,mjsp1,mjsp2,mjsp3,mourik,deng,das,finck} 
and in spintronics for controlling the spin-orbit interaction~\cite{bernevig,kohda} 
have diffused the situation.

A set of three potentials is needed to realize topologically nontrivial superconductors artificially, 
 namely the spin-orbit coupling, the Zeeman field and the pair potential. 
Among them, the spin-orbit interaction mainly affects the energy spectra of the edge states.
In InSb or GaAs, for example, the Dresselhaus spin-orbit
interactions~\cite{drso} are large in films growing along the [110] crystal direction. 
Theoretical studies~\cite{smsc3,flds} have shown that such artificial superconductors 
 host more than one ZES similar to those of the $p_x$-wave superconductor.
We also confirm that a proximitzed spin helix thin film~\cite{bernevig,kohda} also 
traps the flat ZESs with appropriate tuning of the Zeeman field. 
The Dresselhaus superconductors preserve an additional chiral symmetry independently of 
the particle-hole symmetry~\cite{schnyder}. 
Recent theoretical studies~\cite{tewari,niu,diez} have shown that  
the chiral symmetry is responsible for the stability of more than one Majorana fermion. 
On the basis of the above novel insight, we solve an outstanding 
problem regarding the anomalous proximity effect.

In this paper, we first demonstrate the anomalous proximity effect of 
the Dresselhaus superconductors in large magnetic fields.
After discussing the unitary equivalence between the Hamiltonian of Dresselhaus 
and spin-triplet $p_x$-wave superconductors, we analyze the chiral property
of ZESs both at the edge of the superconductor and at the normal 
metal attached to it.
The results show that all the ZESs have the same 
chirality due to the effective $p_x$-wave pairing symmetry.
The pure chiral nature of the wave function 
is responsible for the robustness of highly degenerate ZESs 
in the presence of potential disorder. 
We will prove the perfect Andreev reflection from a $p_x$-wave superconductor 
into a dirty normal metal at zero energy.
This paper provides a microscopic understanding of the anomalous proximity effect
and a design for an artificial $p_x$-wave superconductor.

%***********************************************************************************
\section{Conductance quantization}
%***********************************************************************************
%\textit{Anomalous Proximity Effect}: 

\begin{figure}[hhhh]
\begin{center}
\includegraphics[width=0.48\textwidth]{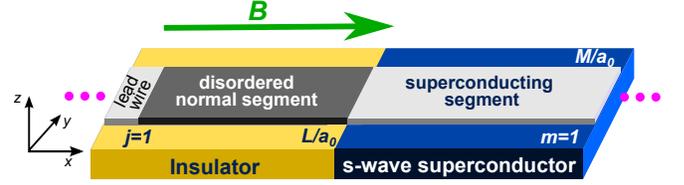}
\caption{(Color online) Schematic image of the NS junction of a Dresselhaus 
superconductor. The superconductor proximitizes 
InSb thin film, which is grown along the [110] crystal direction. 
}
\label{fig:junct}
\end{center}
\end{figure}

\begin{figure}[hhhh]
\begin{tabular}{cc}
 \begin{minipage}{0.22\textwidth}
 \begin{center}
   \includegraphics[width=1.0\textwidth]{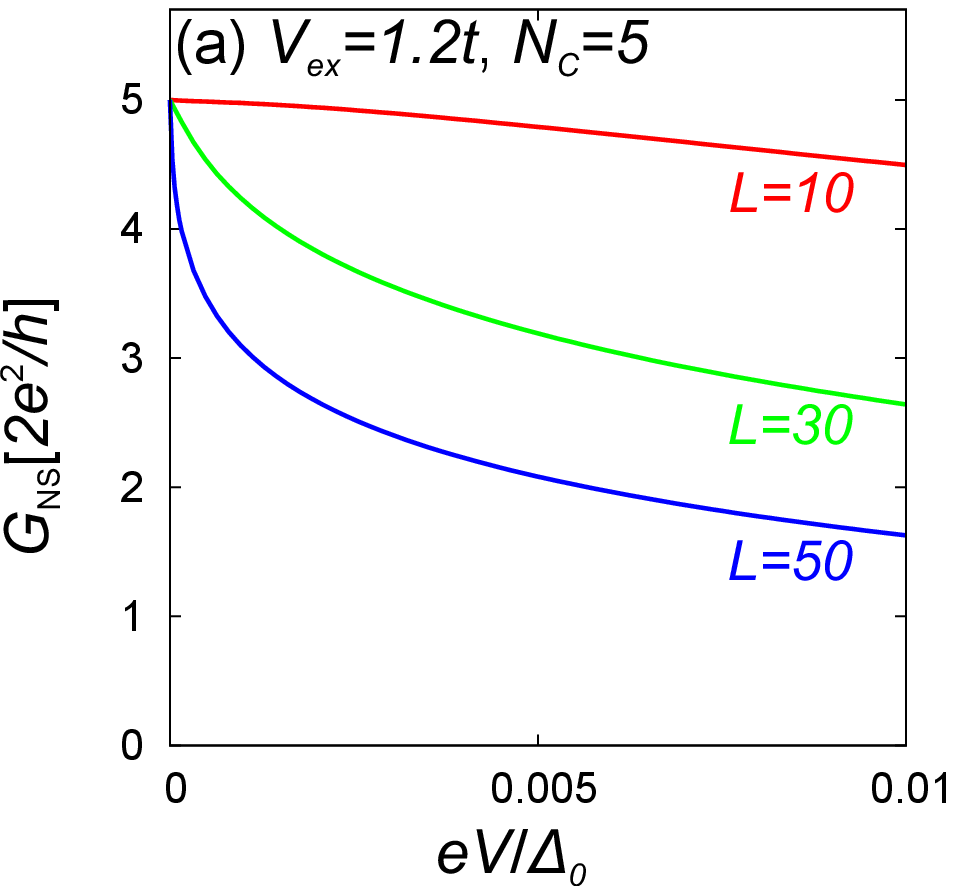}
 \end{center}
 \end{minipage}
 \begin{minipage}{0.22\textwidth}
 \begin{center}
   \includegraphics[width=1.0\textwidth]{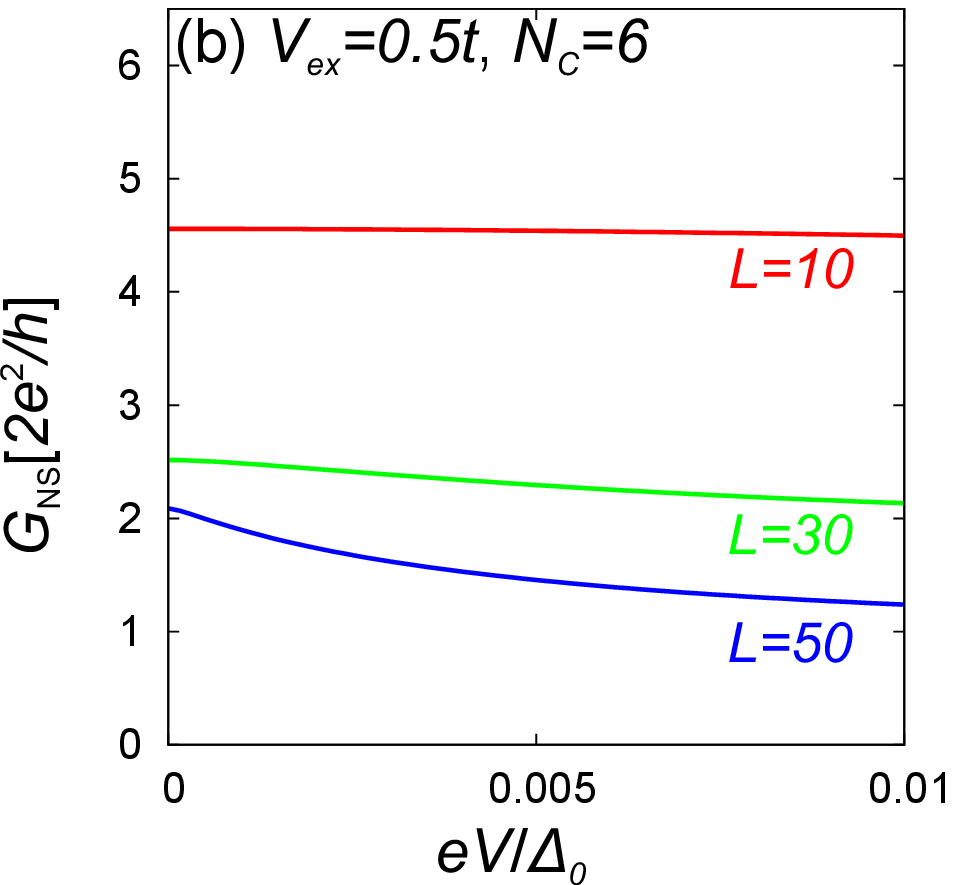}
 \end{center}
 \end{minipage}
\end{tabular}
 \caption{(Color online) The differential conductance is plotted as a function
of the bias voltage for several lengths of disordered 
segment $L$ in units of $a_0$. In (a), the Zeeman potential $V_{ex}=1.2t$ is chosen 
that is larger than a critical value 
of $V_c=0.92t$. 
The number of propagating channels $N_{c}$ is $5$.
In (b), we choose $V_{ex}=0.5t<V_c$ leading to $N_c=6$. 
%The number of samples used for the random ensemble average is $10^{3}.$
}
\label{fig:gns-en}
\end{figure}

First, we numerically demonstrate the anomalous proximity effect of the 
Dresselhaus superconductor. Let us consider an NS junction on 
a two-dimensional tight-binding model with the lattice constant $a_0$ as shown in Fig.~\ref{fig:junct}.
A lattice site is indicated by a vector $\boldsymbol{r}=j\boldsymbol{x}+m\boldsymbol{y}$, 
where $\boldsymbol{x}$ ($\boldsymbol{y}$ )
is the vector in the $x$ ($y$) direction with $|\boldsymbol{x}|= |\boldsymbol{y}|=a_0$.
The present junction consists of three segments:
an ideal lead wire ($-\infty \leq j \leq 0$), a normal disordered segment ($ 1 \leq j \leq L/a_0$)
and a superconducting segment ($L/a_0+1 \leq j \leq \infty$).
The Hamiltonian reads, 
\begin{align}
&\hat{H}_{0} = -t \sum_{\sigma=\uparrow, \downarrow}\sum_{j}\sum_{m=1}^{M/a_0}
\left\{c_{\boldsymbol{r}+\boldsymbol{x},\sigma}^{\dagger} c_{\boldsymbol{r}, \sigma}  +  c_{\boldsymbol{r},\sigma}^{\dagger} c_{\boldsymbol{r}+\boldsymbol{x}, \sigma} \right\} \nonumber\\
&-t \sum_{\sigma=\uparrow, \downarrow} \sum_{j} \sum_{m=1}^{M/a_0-1}
\left\{c_{\boldsymbol{r}+\boldsymbol{y},\sigma}^{\dagger} c_{\boldsymbol{r}, \sigma}  +  c_{\boldsymbol{r},\sigma}^{\dagger} c_{\boldsymbol{r}+\boldsymbol{y}, \sigma} \right\}
\nonumber\\
% c_{\boldsymbol{r},\sigma}^{\dagger}c_{\boldsymbol{r},\sigma} 
&+ \sum_{\boldsymbol{r},\sigma} \left[ 4t-\mu  \right] 
c_{\boldsymbol{r},\sigma}^{\dagger}c_{\boldsymbol{r},\sigma}
+\sum_{j>L/a_0, m} \Delta_{0} \left( 
c_{\boldsymbol{r},\uparrow}^{\dagger} c_{\boldsymbol{r},\downarrow}^{\dagger} + 
H.c. \right)
 \nonumber\\
&- \sum_{\boldsymbol{r},\sigma,\sigma'} 
V_{ex}(\sigma_{1})_{\sigma,\sigma'}
c_{\boldsymbol{r},\sigma}^{\dagger} 
c_{\boldsymbol{r},\sigma'}
+ \sum_{1\leq j\leq L/a_0, m, \sigma} \!\!\! V_{\textrm{imp}}(\boldsymbol{r}) 
c_{\boldsymbol{r},\sigma}^{\dagger}c_{\boldsymbol{r},\sigma} \nonumber
\\
 &- i \frac{\lambda_{D}}{2a_0} \sum_{\boldsymbol{r},\sigma,\sigma'}
(\sigma_{3})_{\sigma,\sigma'} \left(
c_{\boldsymbol{r} + \boldsymbol{x},\sigma}^{\dagger} c_{\boldsymbol{r} ,\sigma'}  -  
c_{\boldsymbol{r},\sigma}^{\dagger} c_{\boldsymbol{r} + \boldsymbol{x},\sigma'} \right),
\label{eq:h0tb}
\end{align}
where $c_{\boldsymbol{r},\sigma}^{\dagger}$($c_{\boldsymbol{r},\sigma}$) is the 
creation (annihilation) operator of an electron
at the site $\boldsymbol{r}$ with spin $\sigma = ( \uparrow$ or $\downarrow )$,
$t=\hbar^2/(2 m a_0^2)$ denotes the hopping integral between the nearest neighbor sites, 
$m$ is the effective mass of an electron,
$\mu$ is the chemical potential, and
 $\lambda_{D}$ represents the strength of the Dresselhaus [110] spin-orbit interaction.
By tuning the magnetic field $B$ in the $x$ direction, 
it is possible to introduce the external Zeeman potential $V_{ex}$.
The parameters $t$, $\mu$, $\lambda_D$ and $V_{ex}$  
are common to the superconductor and the normal metal.
In the $y$ direction, the number of lattice sites is $M/a_0$ and the hard-wall boundary condition 
is applied.
The Pauli matrices in spin space are represented by $\hat{\sigma}_{j}$ for $j = 1-3$
and the unit matrix in spin space is $\hat{\sigma}_{0}$.
We consider the impurity potential given randomly
in the $-W/2 \leq V_{\textrm{imp}}(\boldsymbol{r}) \leq W/2$ range in the normal segment 
($1 \leq j \leq L/a_0$)
and
the $s$-wave pair potential $\Delta_{0}$ in the superconducting segment ($ L/a_0 +1 \leq j \leq \infty$).
%We measure the energy and the length in the units of $t$ and the lattice constant, respectively. 
%----------------------------------------------------
%-----------------------------------------------------
We calculate the differential conductance $G_{\textrm{NS}}$
of the NS junctions based on
the formula\cite{btkf}
\begin{align}
G_{{\rm NS}}(eV) = \frac{e^{2}}{h} \sum_{\zeta,\eta}
\left[ \delta_{\zeta,\eta} - \left| r^{ee}_{\zeta,\eta} \right|^{2}
+ \left| r^{he}_{\zeta,\eta} \right|^{2} \right]_{eV=E},
\end{align}
where $r^{ee}_{\zeta,\eta}$ and $r^{he}_{\zeta,\eta}$ denote
the normal and Andreev reflection coefficients at energy $E$, respectively.
The indices $\zeta$ and $\eta$ label the outgoing and incoming channels, respectively.
These reflection coefficients are obtained
by using the lattice Green's function method\cite{grf1,grf2,fisher}.
With this method, it is possible to calculate the transport coefficients exactly 
even in the presence of random impurity potentials.
In Fig.~\ref{fig:gns-en},
we show the differential conductance of the Dresselhaus superconductors
as a function of the bias voltage for several lengths of disordered 
segments $L$, where we choose parameters of 
 $\mu=1.0t$, $\lambda_{D}=0.2t a_0$, $W=2.0t$, $M=10a_0$ and $\Delta_{0}=0.1t$.
The results are normalized to $G_{Q}=2e^{2}/h$.
In Fig.~\ref{fig:gns-en}(a), we choose $V_{ex}=1.2t$, which leads to the propagating channel number $N_{c}=5$.
The differential conductance decreases with increasing $L$ for finite bias voltages.
However, the zero-bias conductance is quantized at
$G_{Q}N_{c}$ irrespective of $L$.
The results suggest that there are $N_c$ perfect transmission channels  
in a disordered normal segment~\cite{yt04}. 
The conductance quantization at zero bias is an aspect of the 
anomalous proximity effect. 
 We have also confirmed the fractional current-phase 
relationship in SNS junctions~\cite{ya06,dnps4}.
Such anomalous behaviors can be seen when the Zeeman field exceeds
 a critical value $V_{ex}>V_c=\sqrt{\mu_0^2 + \Delta_0^2}$ with 
$\mu_0=\mu-2t\left(1-\cos\left\{\pi/(M/a_0+1)\right\}\right)$.
With the present parameter choice, we obtain $V_c=0.92t$. 
On the other hand for $V_{ex}<V_c$,
the conductance quantization is absent as shown 
in Fig.~\ref{fig:gns-en}(b), where we choose $V_{ex}=0.5t<V_c$.
The zero-bias conductance quantization at 
$G_{Q}N_{c}$ is a robust 
phenomenon in the topologically nontrivial phase described by $V_{ex}>V_c$ and $\lambda_D \neq 0$
and is independent of such parameters as $M$, $W$ and $\mu$.
In the experiment~\cite{mourik}, for instance, the condition $V_{ex}>V_c$
 may be satisfied under a magnetic field of less than 1 tesla in InSb nanowires 
owing to its large g-factor and small Fermi energy. 
Inversion symmetry in the $z$ direction is broken in the junction shown in Fig.~\ref{fig:junct}.
In such case, Rashba spin-orbit interaction $\lambda_R ( k_y \hat{\sigma}_1 - k_x \hat{\sigma}_2 )$
is not negligible. Unfortunately, the Rashba term easily destroys the conductance 
quantization at zero bias because it breaks chiral symmetry discussed below. 
To delete the Rashba term, we need to recover inversion symmetry 
by attaching an appropriate insulator or the same superconductor on top of InSb thin film. 

%---------------------------------------------------
%-----------------------------------------------------------

%***************************************************************
\section{More than one Majorana fermion}
%\label{sec:aatmbs}
%***************************************************************
%\textit{Chiral Symmetry}:  

Secondly, we analyze the chiral property of the ZESs.
In what follows, we consider a Dresselhaus superconductor 
in continuous space for simplicity, 
The Hamiltonian is represented by
\begin{align}
\check{H}_{0} = \left[
\begin{array}{cc}
\hat{h} & i \Delta_{0} \hat{\sigma}_{2} \\
- i \Delta_{0} \hat{\sigma}_{2} & - \hat{h}^{*} \\
\end{array}
\right],
\label{eq:original}
\end{align}
\begin{align}
\hat{h} = \xi_{\boldsymbol{r}} \hat{\sigma}_{0}
- V_{ex} \hat{\sigma}_{1}
+ i \lambda_D \partial_{x} \hat{\sigma}_{3} +V_{\textrm{imp}}(\boldsymbol{r})\hat{\sigma}_0, \label{eq:horg1}
\end{align}
with $\xi_{\boldsymbol{r}} = \frac{- \hbar^{2}}{2m}\nabla^2 -\mu$. 
In the superconductor, 
we consider the uniform pair potential $\Delta_0$ and ignore the impurity potential $V_\mathrm{imp}=0$.
In the normal metal, on the other hand, we introduce the impurity potential and do not consider the pair potential.
We assume a sufficiently large Zeeman potential 
so that $\alpha_D \equiv \lambda_D k_F / V_{ex} \ll 1$ is satisfied 
with $k_F=\sqrt{ 2 m \mu}/\hbar$. 

By applying the unitary transformations as shown in Appendix~\ref{sec:trbdg},
$H_0$ is transformed into $\check{H}_1
= \check{H}_{P} + \check{V}_{\Delta}$ 
within the first order of $\alpha_D$, where 
\begin{align}
\check{H}_{P}
=& \left[
\begin{array}{cc}
\hat{H}_{\uparrow} & 0 \\
0 & \hat{H}_{\downarrow} 
\end{array}
\right],\;  \check{V}_{\Delta}
= \left[
\begin{array}{cc}
0 & i \Delta_{0} \hat{\sigma}_{2} \\
- i \Delta_{0} \hat{\sigma}_{2} &0
\end{array}
\right],
\label{eq:spinmix}
\\
\hat{H}_{\sigma}
=& \left[
\begin{array}{cc}
\xi_{\boldsymbol{r}} +s_s V_{ex} +V_{\textrm{imp}} & -s_s  \frac{\lambda_D \Delta_{0}}{V_{ex}} \partial_{x} \\
s_s \frac{\lambda_D \Delta_{0}}{V_{ex}} \partial_{x} & - \xi_{\boldsymbol{r}} -s_s V_{ex}-V_{\textrm{imp}} 
\end{array}
\right], \label{eq:hpxsigma}
\end{align}
and $s_s=1$ ($-1$) for $\sigma=\uparrow$ ($\downarrow$). 
A Hamiltonian $\hat{H}_{\sigma}$ with $V_{\mathrm{imp}}=0$ 
is equivalent to that of a spin-triplet $p_{x}$-wave 
superconductor and $\check{V}_{\Delta}$ mixes the two spin sectors.
In the topologically nontrivial phase $V_{ex}>V_c$, all the spin-$\uparrow$ states pinch off from the Fermi level 
and only the spin-$\downarrow$ states remain at the Fermi level.
Therefore the spin-mixing term $\check{V}_{\Delta}$ does not affect the remaining 
spin-$\downarrow$ states at all.
In this way, we can shrink the $4\times 4$ Hamiltonian $\check{H}_{1}$ 
of the Dresselhaus superconductor
to the $2\times 2$ Hamiltonian $\hat{H}_{\downarrow}$ of the $p_x$-wave superconductor. 
We assumed the two conditions $\lambda_D k_F \ll V_{ex}$ and $V_{ex} > V_c$ independently.
To realize the $p_x$-wave superconductor, they can be unified into one condition 
$\lambda_D k_F \ll V_c$, which is accessible in the experiment~\cite{mourik}. 
The Hamiltonian $\hat{H}_{\downarrow}$ preserves a chiral symmetry
\begin{align}
\hat{\tau}_1 \;\hat{H}_{\downarrow} \; \hat{\tau}_1 = - \hat{H}_{\downarrow}, 
\quad  \hat{\tau}_1=\left[ \begin{array}{cc} 0 & 1 \\ 1 & 0 \end{array} \right],
\end{align}
where $\hat{\tau}_j$ for $j=1-3$ are the Pauli matrices in Nambu space.
Here we summarize two important 
features of the eigenstates of $\hat{H}_{\downarrow}$ 
proved in Ref.~\onlinecite{chral}. (See also Appendix~\ref{sec:stzs} for details.)

\noindent (i) The eigenstates of $\hat{H}_{\downarrow}$ at zero energy 
are simultaneously the eigenstates of $\hat{\tau}_1$.
Namely, the eigen vectors at zero energy $\varphi_{\nu_0,\lambda}(\boldsymbol{r})$ satisfy
\begin{align}
\hat{H}_{\downarrow} \; \varphi_{\nu_0,\lambda}(\boldsymbol{r})=0, \qquad
\hat{\tau}_1 \; \varphi_{\nu_0,\lambda}(\boldsymbol{r}) = 
\lambda \; \varphi_{\nu_0,\lambda}(\boldsymbol{r}),\label{bdg22}
\end{align}
where $\lambda=\pm 1$ represents the eigen value of $\hat{\tau}_1$ and $\nu_0$ 
is the index of the ZESs.
We have omitted the spin index from the subscripts of $\varphi_{\nu_0,\lambda}$ because 
spin is always $\downarrow$.

\noindent (ii) In contrast to the zero-energy states, the nonzero-energy states 
are not the eigenstates of $\hat{\tau}_1$. They are described by 
the linear combination of two states: one has $\lambda=1$ and the other has $\lambda=-1$. 
%In Fig.~\ref{fig:sta00}, we schematically illustrate the property 
%of the eigen states in the presence of the chiral symmetry. 
Below we prove the robustness of the highly degenerate ZESs in 
a dirty normal segment and the perfect Andreev reflection 
by taking these features into account.

\begin{figure}[hhhh]
\begin{center}
\includegraphics[width=0.4\textwidth]{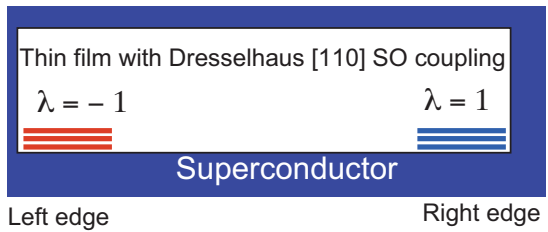}
\caption{(Color online) 
Schematic image of ZESs at two edges of
an isolating Dresselhaus superconductor for $V_{ex}>V_c$.
The number of the ZESs at either edge is equal to $N_c$.
All of the ZESs at the left (right) edge have $\lambda=-1$ ($\lambda=1$).
}
\label{fig:sta00}
\end{center}
\end{figure}

In an isolating Dresselhaus superconductor (i.e., $-L\leq x \leq L$ and $0\leq y \leq M$),
 we can describe 
the wave function of the zero-energy state for each transport channel. 
From the second equation in Eq.~(\ref{bdg22}), it is given by
\begin{align}
\varphi_{n,\lambda}(\boldsymbol{r}) =&
 \chi_{n,\lambda}(x) \; Y_n(y)
\left[
\begin{array}{c}1\\ \lambda \end{array}\right], \label{varphi0}
\end{align}
where $Y_n(y)=\sqrt{2/M}\sin( n \pi y /{M} )$ is the wave function in the $y$ 
direction with the hard-wall boundary condition and $n$ indicates the transport channel.
In the $x$ direction, we also apply the hard-wall boundary condition at its edges,
$
\chi_{n,\lambda}(-L)
= \chi_{n,\lambda}(L)
= 0$.
By substituting Eq.~(\ref{varphi0}) into the first equation in Eq.~(\ref{bdg22}),
we obtain
\begin{align}
\left[ \partial_x^2 - 2 \frac{\lambda}{\xi_D} \partial_{x} +k_n^2
\right] \chi_{n,\lambda}(x) = 0,
\label{eq:eqspin}
\end{align}
where
$\xi_{D} = \xi_0/\alpha_D$, $\xi_0=\hbar v_F/\Delta_0$,
$k_{n} = \sqrt{2m(\mu + V_{ex} -\epsilon_n)} /{\hbar}$, and
$\epsilon_{n} =  
(\hbar n \pi /M )^{2}/(2m)$ is the kinetic energy in the $y$ direction. 
The superconductor must be long enough to satisfy $L/\xi_{D} \gg 1$.
We find the following two solutions for each propagating channel 
\begin{align}
\varphi_{n,-}^{L} (\boldsymbol{r})
=& \frac{C_{L}}{\sqrt{2}}
\left[
\begin{array}{cccc} 1 \\ -1 \\
\end{array}
\right]
{\rm sin}[ q_n (x+L)]
e^{-x/\xi_{D}}\, Y_n(y),\label{eq:leftedg}
\\
\varphi_{n,+}^{R} (\boldsymbol{r})
=& \frac{C_{R}}{\sqrt{2}}
\left[
\begin{array}{cccc}
 1 \\ 1 \\
\end{array}
\right]
{\rm sin}[q_n(x-L)]
e^{x/\xi_{D}}\, Y_n(y),
\label{eq:rightedg}
\end{align}
with $q_n^2= k_n^2-\xi_D^{-2}$, where $C_{L}$ and 
 $C_{R}$ are the normalization coefficients.
We choose the gauge so that the wave functions in Eqs.~(\ref{eq:leftedg}) and (\ref{eq:rightedg})
are real values. 
All the ZESs at the left (right) edge have $\lambda=-1$ ($\lambda=1$), 
which is shown schematically in Fig.~\ref{fig:sta00}.
%*****************************************************************************

In the Bogoliubov transformation, the field operator of an electron with spin-$\downarrow$
is generally described as
\begin{align}
\left[\begin{array}{c}
\Psi (\boldsymbol{r}) \\
\Psi^\dagger(\boldsymbol{r}) 
\end{array}\right]
=&
\sum_{\nu}
\left[ \varphi_{\nu}(\boldsymbol{r}) \gamma_{\nu}
+ \hat{\Xi} \varphi_{\nu}(\boldsymbol{r}) \gamma_{\nu}^{\dagger} \right],
\label{bogoliubov}\\
\hat{\Xi} =& \hat{\tau}_1 {\cal K},\quad  
\hat{\Xi} \hat{H}_{\downarrow} \hat{\Xi}^{-1}= -\hat{H}_{\downarrow}, \label{eq:phsymmetry}
\end{align}
where $\gamma_{\nu}^\dagger$ ($\gamma_{\nu}$) is 
the creation (annihilation) operator
of the Bogoliubov quasiparticle belonging to $E_{\nu}$ and
$\hat{\Xi}$ is the charge conjugation operator
 with ${\cal K}$ indicating a complex conjugation.
Eq.~(\ref{eq:phsymmetry}) represents the particle-hole symmetry of the Hamiltonian.
From Eq.~(\ref{bogoliubov}), we can extract
the electron field operator of the ZES at the $n$-th propagating channel as
\begin{align}
\Psi_{n} (\boldsymbol{r}) =&
i\gamma_{n}^{L} (\boldsymbol{r}) + \gamma_{n}^{R} (\boldsymbol{r}),\\
\gamma_{n}^{L} (\boldsymbol{r}) =& -i
\varphi_{n,-}^{L}(\boldsymbol{r}) 
\left[ \gamma_{n -} - \gamma_{n -}^{\dagger} \right],
\\
\gamma_{n}^{R} (\boldsymbol{r}) =& 
\varphi_{n,+}^{R}(\boldsymbol{r}) 
\left[ \gamma_{n +} + \gamma_{n +}^{\dagger} \right].
\end{align}
The operator $\gamma_{n}^{L} (\boldsymbol{r})$ 
is pure imaginary 
while $\gamma_{n}^{R} (\boldsymbol{r})$ is real in the present gauge choice. 
They satisfy the Majorana relation 
$\left[ \gamma_{n}^{L(R)} (\boldsymbol{r}) \right]^{\dagger} = \gamma_{n}^{L(R)} (\boldsymbol{r})$.
Therefore, 
the number of Majorana fermions at each edge is 
equal to the number of propagating channels at the spin-$\downarrow$ sector $N_\downarrow$.
Since there are no spin-$\uparrow$ channels for $V_{ex}>V_c$, 
 $N_\downarrow$ is equal to $N_c$.
They are degenerate at zero energy at the same place. 
Such highly degenerate states may be fragile in the presence of 
random impurity potential near the edges.
However, at the left edge for example, all of the ZESs have
$\lambda=-1$ as shown in Eq.~(\ref{eq:leftedg}).
According to property (ii), such highly degenerate ZESs are robust 
against potential disorder 
because the random potentials preserve the chiral symmetry
and the ZESs with $\lambda=1$ are absent.

%----------------------------------------------------------
%----------------------------------------------------------

%***************************************************************
\section{Perfect Andreev Reflection and Chiral Nature}

%***************************************************************

Finally and most importantly, we prove the stability of the highly degenerate ZESs 
 in a normal metal.
To analyze the conductance of the NS junction, we attach a normal metal 
to the left side of the superconductor as shown in Fig.~\ref{fig:junct}. 
The Hamiltonian of the normal metal 
is given by $H_{\downarrow}$ in Eq.~(\ref{eq:hpxsigma}) with $\Delta_0=0$.
In the absence of impurity potentials, the wave function in the normal segment at $E=0$ 
is decribed by
\begin{align}
\varphi_N (\boldsymbol{r}) =&  
\sum_n  \left[ 
\left[
\begin{array}{c}  1  \\
 r^{he}_n  \end{array}\right]
 e^{ ik_{n}x}
 +
\left[
\begin{array}{c}  r^{ee}_n \\
 0  \end{array}\right]
 e^{-ik_{n}x}
 \right]
  Y_n(y), \label{eq:wfnormal}
 \end{align}
where 
 $r^{ee}_n$ ($r^{he}_n$) is the normal (Andreev) reflection coefficient 
 at channel $n$ and $k_n=\sqrt{2m(\mu-\epsilon_n)}/\hbar$.
The current conservation law implies 
$\left| r^{ee}_n\right|^2+\left| r^{he}_n\right|^2=1$ 
at $E=0$ for each channel.
From the boundary conditions at the NS interface, the reflection coefficients are calculated to be 
\begin{align}
r^{ee}_n =0, \quad r^{he}_n =-1, \label{reerhe}
\end{align}
for all $n$. 
The wave function in Eq.~(\ref{eq:wfnormal}) is the 
eigenstate of $\hat{\tau}_1$ 
belonging to $\lambda=-1$, (i.e., $\varphi_N \propto [1,-1]^{\textrm{T}}$) 
as well as the ZESs at the left edge of the superconductor. 
According to property (ii), they cannot form nonzero-energy
states. Therefore, the ZESs can penetrate into the normal segment 
while retaining a high degree of degeneracy.
The conclusion is also valid even in the presence of potential disorder
because the impurity potential $V_{\mathrm{imp}}$ preserves the chiral symmetry
and does not damage the pure chiral feature of the ZESs. 
This fact is unique to the $p_x$-wave pairing symmetry.
In the ballistic limit, perfect conductance quantization at zero bias
is a common property of unconventional superconductors that have the edge
ZESs with a flat dispersion. 
For instance, in a spin-singlet 
$d$-wave superconductor with $\Delta_{\boldsymbol{k}} \propto k_x k_y$~\cite{tanaka95}, 
$|r^{he}_{k_y}|=1$ holds for all transverse momentum values $k_y$.
The Hamiltonian of $d$-wave superconductors also preserves the chiral symmetry.
However, the highly degenerate ZESs are fragile under the potential disorder
because the ZESs with two different chiralities coexist at the same edge~\cite{chral}.
Namely, the sign of $r^{he}_{k_y}$ depends on $k_y$.
Therefore the presence of the chiral symmetry is not a sufficient condition 
for the anomalous proximity effect but a necessary one.

In a $p_x$-wave junction, all the ZESs in a normal metal have the same chirality of $\lambda =-1$
in the same way as the ZESs at the left edge of the superconductor. 
The pure chiral feature of the ZESs enables us to explain the 
perfect Andreev reflection into the dirty normal segment. 
According to property (i), the ZESs must be the eigenstate of 
$\hat{\tau}_1$.
We emphasize that the wave function in Eq.~(\ref{eq:wfnormal}) can be the eigenstate 
of $\hat{\tau}_1$ with $\lambda=-1$ only 
when Eq.~(\ref{reerhe}) is satisfied. 
Although the channel index $n$ is no longer a good quantum number 
under the potential disorder, all the wave functions in the normal segment
 have the same vector structure reflecting their pure chiral nature. 
This is a mathematical requirement arising from the chiral symmetry.
The Physical consequence of the vector structure is the 
perfect Andreev reflection into the dirty normal metal at $E=0$.
This explains the perfect quantization of the zero-bias conductance
at $2e^2N_c/h$.

%***************************************************************
\section{Conclusion}
%***************************************************************
%\textit{Conclusion}:
In conclusion, we have discussed the stability of highly degenerate zero-energy states (ZESs)
appearing in disordered junctions consisting of a superconducting 
thin film with Dresselhaus [110] spin-orbit coupling.
The Dresselhaus superconductor hosts more than one ZES at its edges.
When we make a normal-metal/superconductor
junction of the Dresselhaus superconductor, such highly degenerate 
ZESs can penetrate into the dirty normal segment 
and
form resonant transmission channels there. 
An analysis of the wave function in the normal segment shows that 
all the ZESs have the same chirality due to
the effective $p_x$-wave pairing symmetry.
The perfect Andreev reflection into the dirty normal metal  
is a direct consequence of the pure chiral feature of the ZESs.
Our paper provides a microscopic understanding the anomalous 
proximity effect of the spin-triplet $p_x$-wave superconductor.

\begin{acknowledgments}
The authors are grateful to J. D. Sau for useful discussions.
This work was supported by ``Topological Quantum
Phenomena'' (Nos.~22103002, 22103005) and KAKENHI (No.~26287069) from
the Ministry of Education,
Culture, Sports, Science and Technology (MEXT) of Japan
and by the Ministry of Education and Science of the Russian Federation
(Grant No.~14Y.26.31.0007).
\end{acknowledgments}

%***************************************************************
\appendix
%***************************************************************
%**********************************************************************
\section{Zero energy states under a chiral symmetry}
\label{sec:stzs}
%**********************************************************************
Here, we briefly summarize the argument in Ref.~\onlinecite{chral}
which shows the important properties
of zero-energy states under a chiral symmetry.
We consider the BdG Hamiltonian $H$ which preserves the chiral symmetry
\begin{align}
\Gamma H \Gamma^{-1} = - H, \quad 
\Gamma^{2}=1.
\label{eq:c-cs}
\end{align}
The relation is equivalent to
\begin{align}
[ H^2, \Gamma ] = 0.
\label{eq:c-cs2}
\end{align}
The BdG equation is given by
\begin{align}
H \varphi_E (\boldsymbol{r}) = E \varphi_E (\boldsymbol{r}).
\label{eq:c-be}
\end{align}
When we consider the eigen equation of $H^2$,
\begin{align}
H^2 \chi_{E^{2}}(\boldsymbol{r}) = E^{2} \chi_{E^{2}}(\boldsymbol{r}),
\label{eq:chidefa}
\end{align}
Eq.~(\ref{eq:c-cs2}) suggests that the eigen state $\chi_{E^{2}} (\boldsymbol{r})$ 
is also the eigen state of $\Gamma$ at the same time.
Since $\Gamma^{2} = 1$,
we find that the eigen value of $\Gamma$ is $+1$ or $-1$.
Namely the eigen equation
\begin{align}
 \Gamma \chi_{E^2\lambda}(\boldsymbol{r})
= \lambda \chi_{E^2 \lambda}(\boldsymbol{r}),
\label{eq:c-gc}
\end{align}
holds for $\lambda = \pm 1$.
By multiplying $H$ to Eq.~(\ref{eq:c-gc})
from the left side
and by using Eq.~(\ref{eq:c-cs}),
we obtain the equation
\begin{align}
\Gamma H \chi_{E^{2} \lambda}(\boldsymbol{r})
= - \lambda H \chi_{E^{2} \lambda}(\boldsymbol{r}).
\end{align}
We find that $H \chi_{E^{2} \lambda}(\boldsymbol{r})$ is the eigen state of $\Gamma$ 
belonging to $-\lambda$.
Thus we can connect $\chi_{E^{2} +}(\boldsymbol{r})$ and $\chi_{E^{2} -}(\boldsymbol{r})$
as
\begin{align}
H \chi_{E^{2} \lambda}(\boldsymbol{r}) = 
c_{E^{2} \lambda} \chi_{E^{2} -\lambda}(\boldsymbol{r}),
\label{eq:c-chr}
\end{align}
where $c_{E^{2} \lambda}$ is a constant.

The one-to-one correspondence exists between 
$\varphi_E (\boldsymbol{r})$ and
$\chi_{E^2}(\boldsymbol{r})$.
At first, we consider zero-energy states 
$\chi_{0 \lambda}(\boldsymbol{r})$ which satisfies
\begin{align}
H^2 \chi_{0 \lambda}(\boldsymbol{r}) = 0,
\end{align}
in Eq.~(\ref{eq:chidefa}). The integration of $\boldsymbol{r}$ after 
multiplying $\chi_{0 \lambda}^\dagger(\boldsymbol{r})$ from the left
results in
\begin{align}
\int d\boldsymbol{r} \left| H \chi_{0 \lambda}(\boldsymbol{r}) \right|^{2} = 0.
\end{align}
This means that the norm of $H \chi_{0 \lambda}(\boldsymbol{r})$ is zero.
Therefore we conclude that 
\begin{align}
H \chi_{0 \lambda}(\boldsymbol{r})=0. \label{eq:hchizero}
\end{align}
As a result, we find the relation
\begin{align}
\varphi_{0 \lambda}(\boldsymbol{r})
 = \chi_{0 \lambda}(\boldsymbol{r}).
\label{eq:c-pec}
\end{align}
When a zero energy state is described by $\varphi_{0 +}(\boldsymbol{r})
 = \chi_{0 +}(\boldsymbol{r})$, the relations in Eqs.~(\ref{eq:c-chr}) and (\ref{eq:hchizero})
  suggest that $\chi_{0 -}(\boldsymbol{r})=0$.
Therefore the zero-energy states are always the eigen states of $\Gamma$.

For $E \neq 0$, it is possible to represent $\varphi_{E}(\boldsymbol{r})$ 
by $\chi_{E^2 \pm}(\boldsymbol{r})$.
By calculating the norm of $H \chi_{E^{2} \lambda}(\boldsymbol{r})$,
we obtain
\begin{align}
E^{2}=\left| c_{E^{2} \lambda} \right|^{2}.
\end{align}
Multiplying $H$ to Eq. (\ref{eq:c-chr}) from the left alternatively
gives a relation
\begin{align}
c_{E^{2} \lambda}c_{E^{2} -\lambda} = 1.
\end{align}
Therefore,
we find the relation
\begin{align}
H \chi_{E^{2} \lambda}(\boldsymbol{r}) = 
E e^{ i\lambda \theta_{E^{2}}} \chi_{E^{2} -\lambda}(\boldsymbol{r}).
\label{eq:c-chr2}
\end{align}
Although we cannot fix the phase factor $\theta_{E^{2}}$, 
it is possible to express the states $\varphi_E (\boldsymbol{r})$ for $E \neq 0$
as
\begin{align}
\varphi_E (\boldsymbol{r}) =& \frac{1}{\sqrt{2}}
\left( e^{-i\theta_{E^{2}} /2} \chi_{E^{2} +}(\boldsymbol{r})
+s_E  e^{i\theta_{E^{2}} /2}\chi_{E^{2} -}(\boldsymbol{r}) \right),
\label{eq:epm2}
\\
s_E=&\left\{ \begin{array}{cl} 
1 & \text{for}\; E>0 \\
-1 & \text{for}\; E<0.
\end{array}\right.
\end{align}
The nonzero-energy states are constructed
by a pair of eigen states of $\Gamma$: one belongs to $\lambda=1$ and the other belongs $\lambda=-1$.
Therefore, the states with $E \neq 0$ are not the eigen states of $\Gamma$.

%***************************************************************
\section{Unitary Transformation}
\label{sec:trbdg}
%**********************************************************************
The BdG Hamiltonian
of the Dresselhaus nanowire represented by 
\begin{align}
\check{H}_{0} = \left[
\begin{array}{cc}
\hat{h} & i \Delta_{0} \hat{\sigma}_{2} \\
- i \Delta_{0} \hat{\sigma}_{2} & - \hat{h}^{*} \\
\end{array}
\right],
\label{eq:original}
\end{align}
\begin{align}
\hat{h} = \xi_{\boldsymbol{r}} \hat{\sigma}_{0}
- V_{ex} \hat{\sigma}_{1}
+ i \lambda_D \partial_{x} \hat{\sigma}_{3}, \label{eq:horg1}
\end{align}
is transformed 
as follows.
By using the unitary matrix
\begin{align}
\check{R} = \left[
\begin{array}{cc}
\hat{r} & 0 \\
0 & \hat{r}^{*} \\
\end{array}
\right],\;
\hat{r}= \frac{1}{\sqrt{2}}
\left[
\begin{array}{cc}
e^{-i \pi/4} & -e^{-i \pi/4} \\
e^{i \pi/4} & e^{i \pi/4} \\
\end{array}
\right],
\label{eq:unir}
\end{align}
the BdG Hamiltonian $\check{H}_0$ is first transformed to
\begin{align}
\check{H}^{\prime} 
&= \check{R} \check{H}_{0} \check{R}^{\dagger}
\nonumber\\ 
&= \left[
\begin{array}{cc}
\hat{h}^{\prime} & i \Delta_{0} \hat{\sigma}_{2} \\
- i \Delta_{0} \hat{\sigma}_{0} & - \hat{h}^\prime \\
\end{array}
\right],
\label{eq:bdg1}\\
\hat{h}^{\prime} &= \xi_{\boldsymbol{r}} \hat{\sigma}_{0}
+ V_{ex} \hat{\sigma}_{3}
+ i \lambda_D \partial_{x} \hat{\sigma}_{2}.
\end{align}
The Hamiltonian in this basis is represented only by real numbers.
Next we apply a transformation
which is similar to the Foldy-Wouthysen transformation~\cite{fwts}
to the BdG Hamiltonian in Eq.~(\ref{eq:bdg1}). 
Using a unitary matrix
\begin{align}
\check{U} =& \left[
\begin{array}{cc}
\hat{u} & 0 \\
0 & \hat{u} \\
\end{array}
\right],
\label{eq:uniu}\\
\hat{u} =&{\rm exp} [ i \hat{S} ],\;
\hat{S} = \frac{\lambda_D}{2 \hbar V_{ex}} p_{x} \hat{\sigma}_{1},
\end{align}
with $p_{x}= -i \hbar \partial_{x}$,
we transform $H'$ into 
\begin{align}
\check{U} \check{H}^{'} \check{U}^{\dagger}
= \left[
\begin{array}{cc}
e^{i\hat{S}} \hat{h}^{'} e^{-i\hat{S}} & e^{i\hat{S}} (i \Delta_{0} \hat{\sigma}_{2} ) e^{-i\hat{S}}\\
- e^{i\hat{S}} (i \Delta_{0} \hat{\sigma}_{2} ) e^{-i\hat{S}} & - e^{i\hat{S}} \hat{h}^{'} e^{-i\hat{S}} \\
\end{array}
\right].
\end{align}
The diagonal term of Eq.~(\ref{eq:bdg1}) can be expanded as
\begin{align}
e^{i\hat{S}} \hat{h}^{\prime} e^{i\hat{S}} 
= \hat{h}^{\prime} + i [ \hat{S}, \hat{h}^{'} ]
+\frac{i^{2}}{2 !} [ \hat{S}, [ \hat{S}, \hat{h}^{\prime} ] ]
+ \cdots,
\end{align}
with using the Baker-Housdorff formula.
We assume large enough Zeeman potential so that 
$\alpha_D=\lambda_D k_{F}/V_{ex} \ll 1$ is satisfied where
$k_{F} = \sqrt{2m \mu}/\hbar$ denotes Fermi wave number.
From this assumption, we obtain
\begin{align}
e^{i\hat{S}} \hat{h} e^{i\hat{S}}
=\xi \hat{\sigma}_{0}
+ V_{ex} \hat{\sigma}_{3} + O( \alpha_D^2 ),
\end{align}
within the first order of $\alpha_D$.
The off-diagonal term corresponding to the pair potential
is transformed to
\begin{align}
e^{i\hat{S}} (i \Delta_{0} \hat{\sigma}_{2} ) e^{-i\hat{S}}
&= i \Delta_{0} \hat{\sigma}_{2} + i [ \hat{S}, i \Delta_{0} \hat{\sigma}_{2} ] + \cdots
\nonumber\\
&= i \Delta_{0} \hat{\sigma}_{2} - i \frac{\lambda_D \Delta_{0}}{\hbar V_{ex}} p_{x} \hat{\sigma}_{3}
+ O( \alpha_D^2 ),
\end{align}
where we assume the uniform pair potential (i.e., $[p_{x},\Delta_{0}]=0$).
As a result, the BdG Hamiltonian can be written as
\begin{align}
\check{U} \check{H}^{\prime} \check{U}^{\dagger}
= &\left[
\begin{array}{cccc}
\xi_{\boldsymbol{r}} + V_{ex} & 0 & - i \frac{\lambda_D \Delta_{0}}{\hbar V_{ex}} p_{x} & \Delta_{0} \\
0 & \xi_{\boldsymbol{r}} - V_{ex} & - \Delta_{0} &  i \frac{\lambda_D \Delta_{0}}{\hbar V_{ex}} p_{x} \\
i \frac{\lambda_D \Delta_{0}}{\hbar V_{ex}} p_{x} & -\Delta_{0} & -\xi_{\boldsymbol{r}} - V_{ex} & 0 \\
\Delta_{0} &  - i \frac{\lambda_D \Delta_{0}}{\hbar V_{ex}} p_{x} & 0 & -\xi_{\boldsymbol{r}} + V \\
\end{array}
\right] \nonumber\\
&+ O( \alpha_D^2 ).
\end{align}
By interchanging the second column and the third one,
and by interchanging the second row and the third one,
the Hamiltonian can be deformed as
\begin{align}
\check{H}_{1}
=& \check{H}_{P} + \check{V}_{\Delta},
\label{eq:dfbdg}
\\
\check{H}_{P}
=& \left[
\begin{array}{cc}
\hat{H}_{\uparrow} & 0 \\
0 & \hat{H}_{\downarrow} \\
\end{array}
\right],
\\
\hat{H}_{\sigma}
=& \left[
\begin{array}{cc}
\xi_{\boldsymbol{r}} + s_s V_{ex} & - s_s i \frac{\lambda_D \Delta_{0}}{\hbar V_{ex}} p_{x} \\
s_s i \frac{\lambda_D \Delta_{0}}{\hbar V_{ex}} p_{x} & - \xi_{\boldsymbol{r}} -s_s V_{ex} \\
\end{array}
\right],
\\
\check{V}_{\Delta}
=& \left[
\begin{array}{cc}
0 & i \Delta_{0} \hat{\sigma}_{2} \\
- i \Delta_{0} \hat{\sigma}_{2} &0\\
\end{array}
\right],\\
s_s=&\left\{ \begin{array}{cl} 
1 & \text{for}\; \sigma=\uparrow \\
-1 & \text{for}\; \sigma=\downarrow.
\end{array}\right..
\end{align}
These are the starting Hamiltonian in the analytic calculation.

We find that $\check{H}_{1}$ preserves chiral symmetry
\begin{align}
\Gamma  \check{H}_{1} \Gamma^{-1} = - \check{H}_{1},
\quad
\Gamma
= \left[
\begin{array}{cc}
\hat{\sigma}_{1} & 0 \\
0 & \hat{\sigma}_{1} \\
\end{array}
\right].
\end{align}
\par

Finally, we discuss the symmetry property of $H_0$ in Eq.~(\ref{eq:original})
 in its original basis.
It is easy to show that 
 $\check{H}_{0}$ satisfies the relations,
\begin{align}
&\check{\Gamma}_{0} \check{H}_{0} \check{\Gamma}_0^{-1} = - \check{H}_{0}, \quad 
\check{\Gamma}_{0} = \left[ \begin{array}{cc} 0 & -i\hat{\sigma}_1 \\
i\hat{\sigma}_1 & 0 \end{array} \right], \label{eq:chiral4}
\end{align}
which represents the chiral symmetry.
The Hamiltonian $\check{H}_{0}$ also satisfies,
\begin{align}
&\check{\Xi}_{0} \check{H}_{0} \check{\Xi}_0^{-1} = - \check{H}_{0},
\quad
\check{\Xi}_{0} = \left[ \begin{array}{cc} 0 & \mathcal{K}\hat{\sigma}_0 \\
\mathcal{K}\hat{\sigma}_0 & 0 \end{array} \right], \label{eq:charge4}
\end{align}
where $\check{\Xi}_{0}$ represents the charge conjugation 
with $\mathcal{K}$ meaning the complex conjugation.
The first equation in Eq.~(\ref{eq:charge4}) represents the particle-hole 
symmetry.

%**************************************************************************
%\bibliography{apssamp}% Produces the bibliography via BibTeX.

\end{document}